\documentclass[structabstract]{aa}  
\usepackage{graphicx}
\usepackage{txfonts}
\usepackage{natbib}
\bibpunct{(}{)}{;}{a}{}{,} 

\begin{document}
\title{Effects of rotational mixing on the asteroseismic properties of solar-type stars}

\author{P. Eggenberger\inst{1} \and G. Meynet\inst{1} \and A. Maeder\inst{1} \and A. Miglio\inst{2,}\thanks{Postdoctoral Researcher, Fonds de la Recherche Scientifique - FNRS, Belgium} \and J. Montalban\inst{2} \and F. Carrier\inst{3} 
 \and S. Mathis\inst{4,5}  \and C. Charbonnel\inst{1,6} \and S. Talon\inst{7}}

   \offprints{P. Eggenberger}

\institute{Observatoire de Gen\`eve, Universit\'e de Gen\`eve, 51 Ch. des Maillettes, CH-1290 Versoix, Suisse\\ 
	\email{[patrick.eggenberger;georges.meynet;andre.maeder;corinne.charbonnel]@unige.ch}
\and
Institut d'Astrophysique et de G\'eophysique de l'Universit\'e de Li\`ege, 17 All\'ee du 6 Ao\^ut, B-4000 Li\`ege, Belgique\\
         \email{[miglio;montalban]@astro.ulg.ac.be}
\and
Instituut voor Sterrenkunde, Katholieke Universiteit Leuven, Celestijnenlaan 200B, 3001 Leuven, Belgium\\ 
\email{fabien@ster.kuleuven.be}         
\and
Laboratoire AIM, CEA/DSM-CNRS-Universit\'e Paris Diderot, DAPNIA/SAp, 91191 Gif-sur-Yvette Cedex, France\\
\email{stephane.mathis@cea.fr}
\and
LUTH, Observatoire de Paris, CNRS, Universit\'e Paris Diderot, 5 Place Jules Janssen, 92195 Meudon, France
\and
Laboratoire d'Astrophysique de Toulouse-Tarbes, CNRS UMR 5572, Universit\'e Toulouse III, 31400 Toulouse, France 
\and
D\'epartement de Physique, Universit\'e de Montr\'eal, Montr\'eal PQ H3C 3J7, Canada\\
\email{talon@astro.umontreal.ca}
             }

   \date{Received; accepted}

 
  \abstract
   {Observations of solar-like oscillations obtained either from the ground or from space stimulated the study of the effects of various physical processes on the modelling of solar-type stars.}
   {The influence of rotational mixing on the evolution and asteroseismic properties of solar-type stars is studied.}
   {Global and asteroseismic properties of models of solar-type stars computed with and without a comprehensive treatment of shellular rotation are compared. The effects of internal magnetic fields are also discussed in the framework of the Tayler-Spruit dynamo.}
   {Rotational mixing changes the global properties of a solar-type star with a significant increase of the effective
temperature resulting in a shift of the evolutionary track to the blue part of the HR diagram. These differences observed in the HR diagram are related to changes of the chemical composition, because rotational mixing counteracts the effects of atomic
diffusion leading to larger helium surface abundances for rotating models than for non-rotating ones. Higher values of the large frequency separation are then found for rotating models than for non-rotating ones at the same evolutionary stage, because the increase of the effective temperature leads to a smaller radius and hence to an increase of the stellar mean density. In addition to changing the global properties of solar-type stars, 
rotational mixing also has a considerable impact on the structure and chemical composition of the central stellar layers by bringing fresh hydrogen fuel to the central stellar core, thereby enhancing the main-sequence lifetime. The increase of the central hydrogen abundance together with the change of the chemical profiles in the central layers result in a significant increase of the values of the small frequency separations and of the ratio of the small to large separations for models including shellular rotation. This increase is clearly seen for models with the same age sharing the same initial parameters except for the inclusion
of rotation as well as for models with the same global stellar parameters and in particular the same location in the HR diagram. By computing rotating models of solar-type stars including the effects of a dynamo that possibly occurs in the radiative zone, we find that the efficiency of rotational mixing is strongly reduced when the effects of magnetic fields are taken into account, in contrast to what happens in massive stars.}
   {}


   \maketitle
%

\section{Introduction}

The solar five-minute oscillations have provided a wealth of information
on the internal structure of the Sun. These results stimulated various attempts to obtain
similar observations for other solar-type stars. In past years, the spectrographs developed for extra-solar planet search have finally achieved the accuracy needed to detect solar-like oscillations on other stars
\citep[see e.g.][]{bed08}. In addition to these ground-based observations, 
photometric measurements of solar-like oscillations are also obtained from space, thanks to the
CoRoT and the Kepler space missions. All these observations of oscillations for solar-type stars stimulate the theoretical study of the effects of various physical processes on the asteroseismic properties of these stars. Rotation is one of the key processes that influences all outputs of stellar models with a specially strong impact on the physics and evolution of massive stars \citep[see e.g.][]{zah92, mae09}. 
In this work, we study the effects of rotational mixing on the evolution and asteroseismic properties of solar-type stars by comparing stellar models including shellular rotation to non-rotating models. The influence of internal magnetic fields is also discussed in the context of a dynamo that possibly occurs in the radiative zone by computing stellar models including the Tayler-Spruit dynamo \citep{spr02}.  

The modelling of rotation is presented in Sect.~2. The effects of rotational mixing and internal magnetic fields on the evolution and asteroseismic properties of solar-type stars are discussed in Sect.~3, while the conclusion is given in Sect.~4.

\section{Modelling of rotation}

In this section, we briefly summarise the basic physical ingredients of numerical models of rotating stars.  

\subsection{Shellular rotation}

Meridional circulation is generated in the radiative zone of a rotating star as a result of the thermal
imbalance induced by the breaking of the spherical symmetry \citep{edd25, vog26}, structural adjustments and
surface extraction of angular momentum \citep{dec09}.  
Transporting matter and angular momentum, this circulation creates differential rotation 
in the radiative zones, which makes the stellar interior highly turbulent.  
This turbulence is assumed to be much stronger in the horizontal 
than in the vertical direction \citep{tas83, zah92}.
The horizontal turbulent coupling favours an essentially constant angular velocity $\Omega$ on the isobars.
With this hypothesis of shellular rotation, every quantity depends solely
on pressure and can be split into a mean value and its latitudinal perturbation
\begin{equation}
f(P,\theta) = \overline{f}(P) + \tilde{f}(P)P_2(\cos \theta) \, ,
\label{shellular}
\end{equation}
where $P_2(\cos \theta)$ is the second Legendre polynomial.

\subsection{Transport of angular momentum}
\label{trans_mom}

In the framework of shellular rotation, the transport of angular momentum obeys an advection-diffusion 
equation \citep{zah92, mae98}:
\begin{equation}
  \rho \frac{{\rm d}}{{\rm d}t} \left( r^{2}\Omega \right)_{M_r} 
  =  \frac{1}{5r^{2}}\frac{\partial }{\partial r} \left(\rho r^{4}\Omega
  U(r)\right)
  + \frac{1}{r^{2}}\frac{\partial }{\partial r}\left(\rho D r^{4}
  \frac{\partial \Omega}{\partial r} \right) \, , 
\label{transmom}
\end{equation}
where $r$ is the characteristic radius of the isobar, $\rho$ the mean density 
on an isobar, $\Omega(r)$ the mean angular velocity at level $r$ and $D$ is the diffusion
coefficient associated to the transport of angular momentum through turbulent diffusion. In the Geneva stellar evolution code, meridional circulation is treated as a truly advective process. The vertical component $u(r,\theta)$ of the velocity of the meridional circulation at a distance
$r$ to the centre and at a colatitude $\theta$ can be written
\begin{equation}
u(r,\theta)=U(r)P_2(\cos \theta)\,.
\end{equation} 
Only the radial term $U(r)$ appears in Eq. (\ref{transmom});
its expression is given by \cite{mae98}:
\begin{eqnarray}
U(r)  =  \frac{P}{\rho g C_{P} T [\nabla_{\rm ad}-\nabla + (\varphi/\delta)
  \nabla_{\mu}]}
 \times  \left\{ \frac{L}{M}(E_{\Omega }+E_{\mu}) \right\}\,. 
\label{vmer}
\end{eqnarray}
$P$ is the pressure, $C_P$ the specific heat, 
$\nabla = \frac{\partial \ln T}{\partial \ln P}$, 
$\delta = - \left( \frac{\partial \ln \rho}{\partial \ln T} \right)_{P, \mu}$ and
$\varphi = \left( \frac{\partial \ln \rho}{\partial \ln \mu} \right)_{P, T}$ with $\mu$ the mean molecular weight. 
$E_{\Omega}$ and $E_{\mu}$ are terms depending on the $\Omega$- and $\mu$-distributions respectively, up to the third
order derivatives and on various thermodynamic quantities \cite[see][ for more details]{mae98}.

Meridional circulation and shear mixing are considered as the main mixing mechanisms in radiative zones.
The first term on the right-hand side of Eq. (\ref{transmom}) describes the advection of angular momentum by 
meridional circulation, while the second term accounts for the transport of angular momentum by shear instability
with $D=D_{\rm shear}$. The expression of this diffusion coefficient is given by
\begin{eqnarray}
D_{\rm shear} & = & \frac{ 4(K_T + D_{\mathrm{h}})}
{\left[\frac{\varphi}{\delta} 
\nabla_{\mu}(1+\frac{K}{D_{\mathrm{h}}})+ (\nabla_{\mathrm{ad}}
-\nabla_{\mathrm{rad}}) \right] } \nonumber\\
&  & \times \frac{H_{\mathrm{p}}}{g \delta} \; 
 \frac{\alpha}{4}\left( 0.8836 \, \Omega{{\rm d}\ln \Omega \over {\rm d}\ln r} \right)^2 \, ,
\label{dshear}
\end{eqnarray}
with $K_T$ the thermal diffusivity \citep{mae01}. 
$D_{\rm h}$ is the diffusion coefficient associated
to horizontal turbulence. This coefficient can be obtained
by expressing the balance between the energy dissipated by the 
horizontal turbulence and the excess of energy present in the differential rotation \citep{mae03}:
\begin{equation}
D_{\rm h} = A r \left( r {\bar \Omega(r)} V [2V-\alpha U]\right)^{\frac{1}{3}} \,,
\label{Dhmaeder}
\end{equation}
with
\begin{equation}
A= \left( \frac{3}{400 n \pi} \right)^{\frac{1}{3}} \,.
\label{Amaeder}
\end{equation}
$U$ is the vertical component of the meridional circulation velocity, $V$ the
horizontal component, and $\alpha=\frac{1}{2} \frac{{\rm d} \ln r^2 {\bar \Omega}}{{\rm d} \ln r}$. 

The full solution of Eq. (\ref{transmom}) taking into account $U(r)$ and $D$ gives the 
non-stationary solution of the problem. The expression of $U(r)$ (Eq. \ref{vmer}) involves
derivatives up to the third order; Eq. (\ref{transmom}) is thus of the fourth order and implies
four boundary conditions. The first boundary conditions impose momentum conservation at convective 
boundaries: 
\begin{displaymath}
\begin{array}{ll}
{\displaystyle \frac{\partial}{\partial t}
\left[\Omega \int _{r_{\rm t}}^R  r^4 \rho \, {\rm d} r \right] =
-\frac{1}{5} r^4 \rho \Omega U + {\cal F}_\Omega } & ~~~{\rm for} ~ r=r_{\rm t} \vspace{0.2cm}\\
{\displaystyle \frac{\partial}{\partial t}
\left[\Omega \int _0^{r_{\rm b}}  r^4 \rho \, {\rm d}r \right] = \frac{1}{5} r^4 \rho \Omega
U } & ~~~{\rm for} ~~r=r_{\rm b}\,.
\end{array}
\end{displaymath}
The other conditions are determined by requiring the absence of differential rotation at
convective boundaries \citep{tal97}
\begin{equation}
\frac{\partial \Omega}{\partial r} = 0 \; {\rm for} \; r = r_{\rm t},\, r_{\rm b} \, .
\end{equation}
$r_{\rm t}$ and $r_{\rm b}$ correspond respectively to the top (surface) and bottom (center) of the radiative
zone.
${\cal F}_\Omega$ represents the torque
applied at the surface of the star. For solar-type stars, this torque corresponds to the magnetic coupling
at the stellar surface. Indeed, these stars are assumed to undergo magnetic braking during their evolution on the main
sequence. We adopt the braking law of \cite{kaw88}:
\begin{eqnarray}
\nonumber
 \frac{{\rm d} J}{{\rm d}t} = \left\{
\begin{array}{l l }
-K \Omega^3 \left({\displaystyle \frac{R}{R_\odot}} \right)^{1/2} 
\left({\displaystyle \frac{M}{M_\odot} }\right)^{-1/2} & 
(\Omega \leq \Omega_{\rm sat}) \\
\nonumber
 & \\
\nonumber
-K \Omega \, {\Omega^2}_{\rm sat} \left({\displaystyle \frac{R}{R_\odot}} \right)^{1/2} 
\left({\displaystyle \frac{M}{M_\odot} }\right)^{-1/2}
& (\Omega > \Omega_{\rm sat}) \, .  
 \end{array}   \right.
\end{eqnarray}
The constant $K$ is related to the magnitude of the magnetic field
strength; it is usually calibrated on the Sun and taken to be a constant in all
stars \cite[e.g.][]{bou97}. $\Omega_{\rm sat}$ expresses that magnetic field generation saturates at some critical
value \cite[][ and references therein]{saa96}.
This saturation is required in order to retain a sufficient amount of
fast rotators in young clusters, as originally suggested by \cite{sta87}. 
For the present computations, $\Omega_{\rm sat}$ is fixed to 14\,$\Omega_{\odot}$ \cite[see][]{bou97}.

\subsection{Transport of chemical elements}

The vertical transport of chemicals
through the combined action of vertical advection and strong
horizontal diffusion 
can be described as a pure diffusive process \citep{cha92}.
The advective transport is then replaced by a diffusive term, 
with an effective
diffusion coefficient
\begin{equation}
D_{\rm eff} = \frac{|rU(r)|^2}{30D_{\rm h}}\,,
\label{Deff}
\end{equation}
where $D_{\rm h}$ is the diffusion coefficient associated
to horizontal turbulence (Eq. \ref{Dhmaeder}).
The vertical transport of chemical elements then
obeys a diffusion equation which, in addition to this macroscopic transport,
also accounts for (vertical) turbulent transport with the same coefficient 
$D_{\rm shear}$ as for the transport of angular momentum (Eq. \ref{dshear}), nuclear reactions,
and atomic diffusion.

\section{Results}

The stellar evolution code used for these computations is the Geneva code, 
which includes a comprehensive treatment of shellular rotation as briefly described 
in the preceding section \citep[see][ for more details]{egg08}. 
In addition to shellular rotation, atomic diffusion can be included in the computation using the
routines developed for the Toulouse-Geneva version of the code
\citep[e.g.][]{ric96}. The diffusion coefficients are computed according to the prescription by
\cite{paq86}. Diffusion due to concentration and thermal
gradients is included, while the radiative acceleration is neglected since it is negligible for the
structure of low-mass stellar models with extended convective envelopes \citep{tur98}.

\subsection{Main-sequence evolution of rotating models}
\label{sec_evo}

To investigate the effects of rotational mixing on the properties of solar-type stars, the main-sequence evolution of 1\,M$_{\odot}$ stars  
is computed with a solar chemical composition as given by \cite{gre93} and
a solar calibrated value for the mixing-length parameter. 
The main-sequence evolution of non-rotating models with and without atomic diffusion 
of helium and heavy elements is first computed.
Two additional rotating models with an initial velocity of $50$\,km\,s$^{-1}$ on the zero age main sequence
(ZAMS) are computed with and without the inclusion of atomic diffusion of helium and heavy elements.
For these models, the braking constant $K$ is 
calibrated so that a 1\,M$_{\odot}$ star with an initial velocity of $50$\,km\,s$^{-1}$ on the ZAMS
reproduces the solar surface rotational velocity after $4.57$\,Gyr.
All four models share exactly the same initial parameters except for the inclusion of shellular rotation 
and atomic diffusion. 

\begin{figure}[htb!]
 \resizebox{\hsize}{!}{\includegraphics{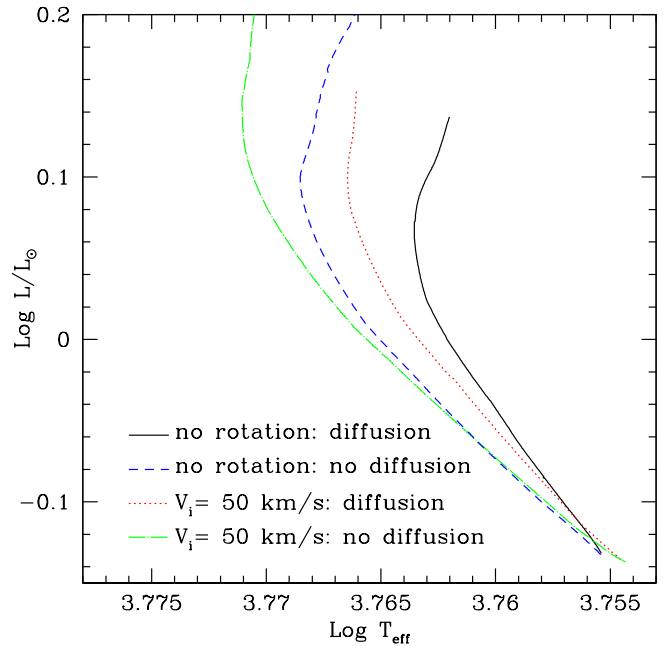}}
  \caption{Evolutionary tracks in the HR diagram for 1\,M$_\odot$ models. The continuous and dashed lines correspond to non-rotating models computed with and without atomic diffusion, respectively.  The dotted and dot-dashed lines indicate rotating model with an initial velocity of 50\,km\,s$^{-1}$ computed with and without atomic diffusion, respectively. The tracks stop at the end of the main sequence.}
  \label{dhr_rotdiff}
\end{figure}

Figure~\ref{dhr_rotdiff} shows the main-sequence evolution in the HR diagram for these models. Comparing rotating
and non-rotating models, we see that the inclusion of rotation changes the evolutionary tracks in the HR diagram.
Rotating models indeed exhibit higher effective temperatures and slightly higher luminosities than non-rotating ones; 
this results in a shift of the evolutionary tracks to the blue part of the HR diagram when rotation 
is taken into account. 
Figure~\ref{dhr_rotdiff} also shows 
that models including atomic diffusion exhibit lower effective temperatures and luminosities than the corresponding
models without atomic diffusion. Contrary to rotation, the inclusion of atomic diffusion is thus found to shift 
evolutionary tracks towards the red part of the HR diagram. 

\begin{figure}[htb!]
 \resizebox{\hsize}{!}{\includegraphics{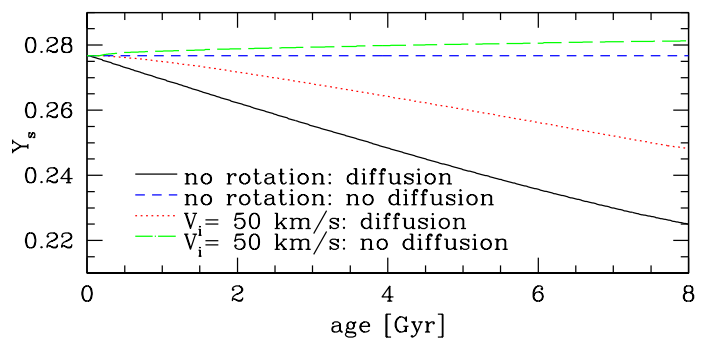}}
  \caption{Surface abundance of helium $Y_{\rm s}$ during the main-sequence evolution for the 1\,M$_\odot$ models shown in Fig.~\ref{dhr_rotdiff}.}
  \label{yst_rotdiff}
\end{figure}

In Fig.~\ref{yst_rotdiff}, the variation of the helium surface abundance $Y_{\rm s}$ during the main sequence is plotted
for the four models shown in Fig.~\ref{dhr_rotdiff}. For the model computed without atomic diffusion and rotation
(dashed line), the helium surface abundance remains constant because no mixing mechanisms are taken into account in the
radiative zone. The inclusion of atomic diffusion leads to a decrease of the helium mass fraction at the stellar surface.
Comparing models including atomic diffusion and computed with and without rotation (dotted and continuous lines), 
a lower decrease of the helium surface abundance is found for the rotating model than for the non-rotating one. We thus
see that rotational mixing counteracts the effects of atomic diffusion in the external layers of the star. As a result,
rotating models exhibit larger surface abundances of helium leading to a decrease of the opacity in the external layers
and hence to the shift towards the blue part of the HR diagram observed in Fig.~\ref{dhr_rotdiff}. 
These differences in the helium content of the external layers of rotating and non-rotating stars 
increase during the main-sequence evolution resulting in significant differences in the HR diagram.

\begin{figure}[htb!]
 \resizebox{\hsize}{!}{\includegraphics{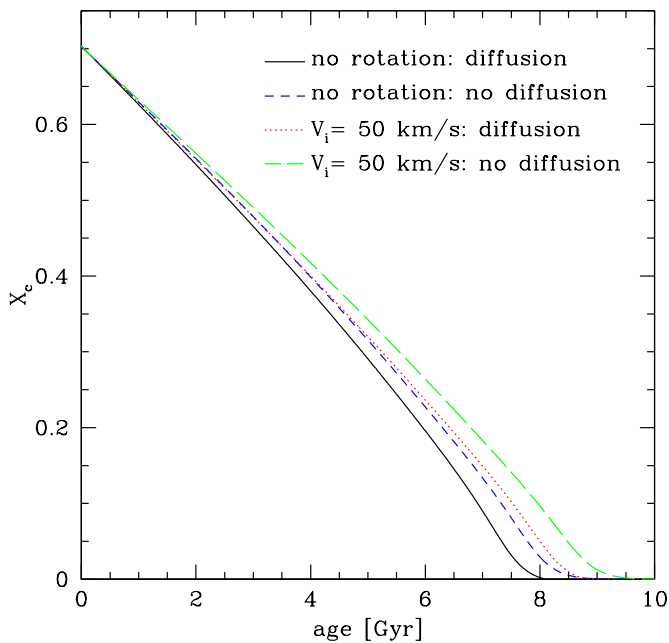}}
  \caption{Same as Fig.~\ref{yst_rotdiff} but for the central hydrogen abundance $X_{\rm c}$. } 
   \label{xct_rotdiff}
\end{figure}

In addition to changing the evolutionary tracks in the HR diagram through the change of 
the surface chemical composition, rotation has also a significant impact on the properties
of the central layers. In particular, rotational mixing brings fresh hydrogen fuel to the central stellar core, 
thereby enhancing the main-sequence lifetime. Figure~\ref{xct_rotdiff} shows the evolution of 
the central mass fraction of hydrogen (X$_{\rm c}$) as a function of the age for the same four models
of 1\,M$_\odot$. Owing to rotational mixing, we see that at a given age the central hydrogen mass fraction 
is larger for rotating models than for models without rotation. On the contrary, the inclusion of atomic
diffusion leads to a more rapid decrease of the central hydrogen abundance. By comparing the model 
including atomic diffusion and rotation (dotted line, model R-D hereafter) to the model computed without atomic diffusion nor rotation 
(dashed line, model NR-ND hereafter), we see that rotational mixing completely counteracts the effects of atomic diffusion 
in the central layers and even leads to values of X$_{\rm c}$ which are slightly higher than for the model
computed without atomic diffusion and rotation. We thus find that the efficiency of rotational mixing relative to
atomic diffusion is higher in the central layers of a solar-type star than in its external layers. 
Figure~\ref{yst_rotdiff} indeed shows that rotation only reduces the efficiency of atomic diffusion 
in the external layers, but does not completely counteract these effects, since the R-D model (dotted line) exhibits
surface abundances of helium that are higher than the values of the non-rotating model including atomic
diffusion (continuous line, model NR-D hereafter) but also lower than the helium surface abundances of the NR-ND model (dashed line).    
As a result of rotational mixing, the main-sequence lifetime is longer for stellar models including rotation. 
For instance, models of 1\,M$_\odot$ computed with an initial velocity of 50\,km\,s$^{-1}$ exhibit 
ages at the end of the main sequence about 10\% higher than the corresponding non-rotating models.

\subsection{Asteroseismic properties of rotating models}

\subsubsection{Models with the same initial parameters}
\label{sec_modini} 

In the preceding section, the effects of rotation on the structure and evolution of solar-type have been discussed.
We are now interested in investigating how these rotational effects change the asteroseismic properties of solar-type
stars. For this purpose, the NR-ND model is first compared to the 1\,M$_\odot$ model computed with an initial velocity of $50$\,km\,s$^{-1}$ and without atomic diffusion (model R-ND herafter).
Both models therefore share the same initial parameters except for the inclusion of shellular rotation. 
The theoretical low-$\ell$ frequencies of these models are then computed by using the Aarhus adiabatic 
pulsation code \citep{chr08}.

\begin{figure}[htb!]
 \resizebox{\hsize}{!}{\includegraphics{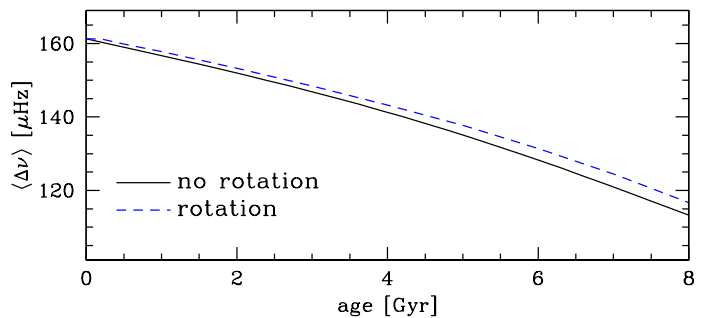}}
  \caption{Mean large separation $\langle \Delta \nu \rangle$ versus age for 1\,M$_\odot$ models computed with
the same initial parameters except for the inclusion of rotation. The rotating model has an initial velocity
of $50$\,km\,s$^{-1}$ on the ZAMS.}
  \label{gde_rotdiff}
\end{figure}

The first asteroseismic quantity we consider is the mean large separation $\langle \Delta \nu \rangle$.
This frequency spacing is defined as the differences 
between oscillation modes with the same angular degree $\ell$ and consecutive radial order $n$: $\Delta \nu_{\ell}(n) \equiv \nu_{n,\ell} - \nu_{n-1,\ell}$.
The mean value of the large separation is calculated from modes
with a radial order $n$ ranging from 15 to 25 for $\ell =0$ and $\ell =1$ modes, and for
an order $n$ ranging from 14 to 24 for $\ell =2$ modes (this interval typically corresponds 
to oscillation modes detected in main-sequence solar-type stars). 
The variation of the mean large separation during the main-sequence evolution
is shown in Fig.~\ref{gde_rotdiff} for the rotating and the non-rotating 1\,M$_\odot$ model. 

\begin{figure}[htb!]
 \resizebox{\hsize}{!}{\includegraphics{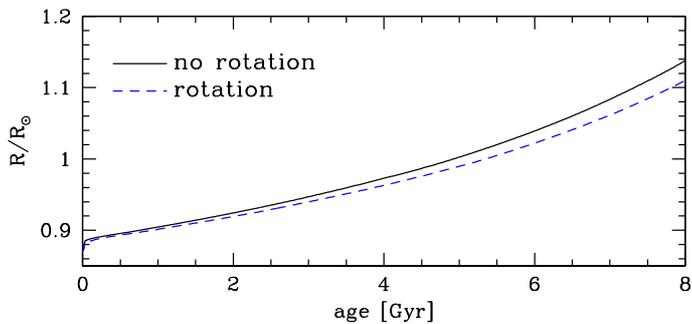}}
  \caption{Variation of the stellar radius during the main-sequence evolution of 1\,M$_\odot$ models 
with and without rotation. The rotating model is computed with an initial velocity of $50$\,km\,s$^{-1}$ on the ZAMS.}
  \label{R_vs_t}
\end{figure}

Figure~\ref{gde_rotdiff} first shows that the mean large separation decreases 
during the evolution of the star on the main sequence. This can be readily understood
by recalling that the large separation is asymptotically proportional to the square root of the star's mean density.
The stellar radius increases during the evolution of the star on the main sequence (see Fig.~\ref{R_vs_t})
resulting thereby in a decrease of the star's mean density and hence of the mean large separation.

Concerning the effects of rotation, Fig.~\ref{gde_rotdiff} shows that rotational mixing increases  
the value of the mean large separation at a given age. The decrease of the mean large separation during the main sequence
is then found to be larger for models without rotation than for rotating models. Frequency differences of 
1.3, 2.3, 3.1 and 3.7\,$\mu$Hz between the values of the mean large separation for a model with an initial
velocity of $50$\,km\,s$^{-1}$ and a non-rotating model 
are reached after 2, 4, 6 and 8\,Gyr, respectively. Since the typical accuracy on the observed mean large separation
obtained so far by ground-based and space observations of solar-like oscillations is better than about 0.5\,$\mu$Hz 
\cite[e.g.][]{cha10,mos09,car06,bed06,car05_bvir,bou02}, we see that the inclusion of rotation leads to significant changes of the mean large separation.  
These effects of shellular rotation on the mean large separation are related to the decrease of
the stellar radius when the rotational velocity increases. Figure~\ref{R_vs_t} indeed shows
that the stellar radius at a given age is larger for the non-rotating model than for the
model including shellular rotation. At a given luminosity, a rotating model is 
characterized by a smaller radius, a higher mean density and hence a higher value of the 
mean large separation than a non-rotating model. 

The effects of rotational mixing on the helium surface abundances can also be directly revealed by asteroseismic
observations. The depression in the adiabatic index $\Gamma_1$ in the second helium ionization zone induces an oscillatory signal in the frequencies of seismic oscillations that can be used to determine the helium abundance 
in the convective envelope of solar-type stars \citep[e.g.][]{mon98,gou02, mig03, bas04,hou07}. Such a determination requires
high accuracy asteroseismic observations. According to \cite{bas04}, estimated errors on the helium abundance $Y$ range from
0.03 for 0.8\,M$_{\odot}$ stars to 0.01 for 1.2\,M$_{\odot}$ stars with a value of about 0.02 for 1\,M$_{\odot}$ stars assuming frequency errors typical of current asteroseismic space missions of one part in $10^4$. Comparing the surface helium abundance of 1\,M$_{\odot}$ models including atomic diffusion and computed with and without rotation (continuous and dotted lines in Fig. \ref{yst_rotdiff}), we see that the difference in the helium abundance of the rotating and non-rotating model increases during the main-sequence evolution and exceeds 0.02 after about 5.5\,Gyr. We thus conclude that such a direct asteroseismic observation of the effects of rotational mixing on the surface helium abundance is of course very difficult to obtain but could be feasible for solar-type stars near the end of the main sequence.

\begin{figure}[htb!]
 \resizebox{\hsize}{!}{\includegraphics{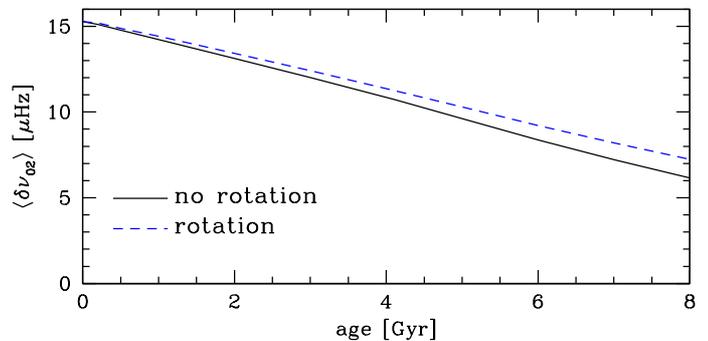}}
  \caption{Same as Fig.~\ref{gde_rotdiff} but for the mean small separation $\langle \delta \nu_{02} \rangle$.}
  \label{pte_rotdiff}
\end{figure}

After the effects of shellular rotation on the global stellar parameters and external stellar layers, 
we now discuss the changes of the structure of the central layers
due to rotational mixing by computing the mean small separations for the rotating and non-rotating
models studied before. The small separation $\delta \nu_{\ell,\ell+2}(n) \equiv \nu_{n,\ell}- \nu_{n-1,\ell+2}$ 
is the difference between the frequencies of modes with an angular degree $\ell$ of
same parity and with consecutive radial order. The mean small separation $\langle \delta \nu_{02} \rangle$
is calculated from $\ell =0$ modes
with a radial order $n$ ranging from 15 to 25 and $\ell =2$ modes with an order $n$ between
14 and 24. The variation of the mean small separation $\langle \delta \nu_{02} \rangle$ 
during the main sequence is shown in Fig.~\ref{pte_rotdiff} for both models.

\begin{figure}[htb!]
 \resizebox{\hsize}{!}{\includegraphics{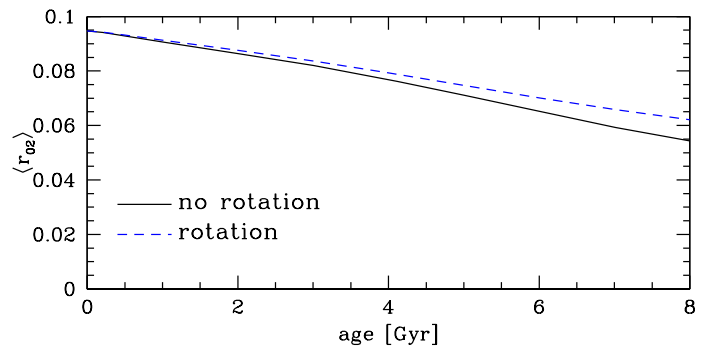}}
  \caption{Same as Fig.~\ref{gde_rotdiff} but for the mean value of the ratio $r_{02}$.}
  \label{r02moy_rotdiff}
\end{figure}

The small separation is very sensitive to the structure of the core and is mainly proportional to
the hydrogen content of the central layers. Figure~\ref{pte_rotdiff} shows
the decrease of $\langle \delta \nu_{02} \rangle$ with the age resulting from the decrease
of the hydrogen abundance in the central core due to nuclear burning.
The inclusion of shellular rotation leads to higher values 
of the mean small separations at a given age. Frequency differences of 
0.3, 0.5, 0.8 and 1.1\,$\mu$Hz between the values of $\langle \delta \nu_{02} \rangle$ for a model with an initial
velocity of $50$\,km\,s$^{-1}$ and a non-rotating model are observed. This increase of the small separation for rotating
models is mainly due to the increase of the central hydrogen abundance by rotational mixing discussed in the
preceding section (see Fig.~\ref{xct_rotdiff}). While the small separation is principally 
sensitive to the conditions in the central regions of the star, it can also retain some sensitivity
to the near-surface structure. \cite{rox03} introduce therefore the use of another asteroseismic
diagnostic: the ratio of small
to large separations $r_{02}(n) \equiv \delta \nu_{02}(n)/ \Delta \nu_{\ell=1}(n)$. This ratio 
constitutes a more reliable diagnostic for the central parts of a star than the small separation,
since it is essentially independent of the structure of the outer layers, and is determined solely
by the interior structure \citep{rox03,flo05}. The variation of the mean ratio $\langle r_{02} \rangle $ with the age is 
shown in Fig.~\ref{r02moy_rotdiff} for both models.

Figure~\ref{r02moy_rotdiff} confirms that the increase of the small separation by rotational mixing observed 
in Fig.~\ref{pte_rotdiff} is mainly due to the rotation-induced changes of the conditions in the central regions 
and not to the change of the global parameters nor of the near-surface structure of the star. Near the
end of the main sequence, an increase of 15\% of the ratio $\langle r_{02} \rangle $ is indeed found when rotational
effects are taken into account. By comparing this value to the corresponding increase of 18\% found for the small separation 
$\langle \delta \nu_{02} \rangle$, we see that the sensitivity of $\langle \delta \nu_{02} \rangle$ 
against changes in the global parameters and near-surface structure induced by rotation (resulting in
an increase of the mean large separation of 3\% near the end of the main sequence) is quite limited.

\begin{table}
\caption{Asteroseismic properties for 1\,M$_{\odot}$ models with an age of 6\,Gyr.}

\begin{center}
\label{tab:6gyr}
\begin{tabular}{l|cc|cc}
\hline
\hline 
Model & atomic & rotation & $\langle \Delta \nu \rangle $ & $\langle r_{02} \rangle $ \\
 & diffusion & & [$\mu$Hz] &  \\ \hline
NR-D & yes & no & $126.3$ & $0.057$  \\
R-D & yes & $V_{\rm ini} = 50$\,km\,s$^{-1}$ & $130.8$ & $0.066$  \\
NR-ND & no & no & $131.3$ & $0.060$  \\
R-ND & no & $V_{\rm ini} = 50$\,km\,s$^{-1}$ & $134.4$ & $0.069$  \\
\hline
\end{tabular}
\end{center}
\end{table}

\begin{figure}[htb!]
 \resizebox{\hsize}{!}{\includegraphics{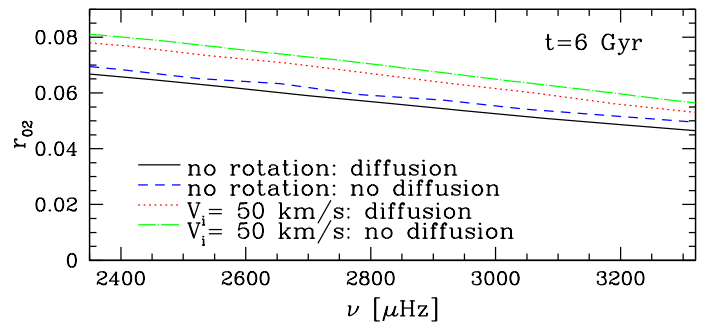}}
  \caption{Variation of the ratio $r_{02}$ with frequency for the 1\,M$_\odot$ models listed in
Table~\ref{tab:6gyr}.}
  \label{r02_rotdiff_t6}
\end{figure}

\begin{figure}[htb!]
 \resizebox{\hsize}{!}{\includegraphics{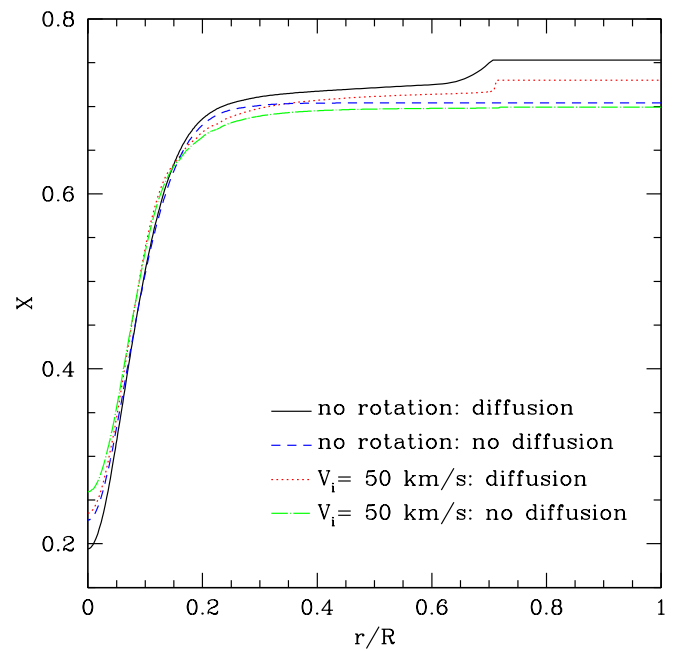}}
  \caption{Hydrogen abundance profiles for the 1\,M$_\odot$ models listed in Table~\ref{tab:6gyr}.}
  \label{profx}
\end{figure}

As mentioned in Sect.~\ref{sec_evo}, the efficiency of rotational mixing relative to
atomic diffusion is found to be greater in the central layers of a solar-type star than in its external layers. 
In order to find out how this difference can be seen in the asteroseismic properties of a star, we now compare 
four different stellar models with the same age but computed with/without atomic diffusion and rotation. The
mean values of the large separation and of the ratio $r_{02}$ are given in Table~\ref{tab:6gyr} for models
with an age of 6\,Gyr. Table~\ref{tab:6gyr} shows that models computed without atomic diffusion exhibit
the higher values of the mean large separation, while rotating models exhibit the higher values of the
mean ratio $r_{02}$. We thus see that rotational effects have a more substantial impact on the ratio $\langle r_{02} \rangle $ than
on the mean large separation $\langle \Delta \nu \rangle $. Rotational mixing is found to only partially 
inhibit the effects of atomic diffusion on the large separation, while the impact of rotation on the 
ratio $r_{02}$ is much more important than the effects of atomic diffusion.
This can be seen by comparing the rotating model R-D to the non-rotating models NR-D and NR-ND. Model R-D indeed 
exhibits a higher value of $\langle \Delta \nu \rangle $ than model NR-D, but a slightly lower value than
model NR-ND, which is computed without atomic diffusion. The situation is quite different for the mean ratio 
$\langle r_{02} \rangle $, because this value is significantly higher for model R-D than for both non-rotating models.
This can also clearly be seen in Fig.~\ref{r02_rotdiff_t6} which shows the variation of $r_{02}$with frequency for
the four models listed in Table~\ref{tab:6gyr}. Recalling that the large separation is principally sensitive 
to the mean stellar density and the ratio $r_{02}$ to the central properties of the star, we conclude
that the greater efficiency of rotational mixing relative to atomic diffusion in the central stellar layers 
is clearly visible in the asteroseismic properties of the models. The increase of $r_{02}$ for
rotating models shown in Fig.~\ref{r02_rotdiff_t6} is of course due to the higher value of $X_{\rm c}$ 
at a given age when rotational effects are taken into account (see Fig.~\ref{xct_rotdiff}), but also to the
changes of the chemical profiles in the central stellar layers due to rotational mixing 
(see Fig.~\ref{profx}). 

\begin{figure}[htb!]
 \resizebox{\hsize}{!}{\includegraphics{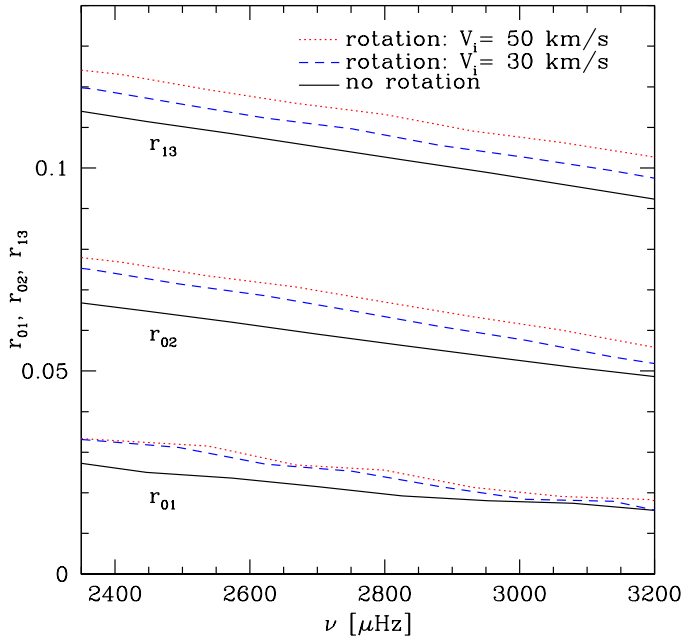}}
  \caption{Ratio $r_{01}$, $r_{02}$ and $r_{13}$ for 1\,M$_\odot$ models with the same age of 6\,Gyr. The continuous line corresponds to a non-rotating model, while the dashed and dotted lines indicate models computed with an initial velocity of $30$\,km\,s$^{-1}$ and $50$\,km\,s$^{-1}$, respectively.}
  \label{r01r02r13_t6}
\end{figure}

The effects of rotational mixing on the central stellar layers have been studied by discussing the changes of the small separation between modes with $\ell=0$ and 
$\ell=2$ and of the corresponding ratio of the small to large separation $r_{02}$. The increase of the small separation and of the ratio of the small to large separation due to rotational mixing is also clearly visible for the separation between $\ell=1$ and  $\ell=3$ modes ($\delta \nu_{13}(n) \equiv  \nu_{n,\ell=1}- \nu_{n-1,\ell=3}$) and between $\ell=0$ and $\ell=1$ modes ($\delta \nu_{01}(n) \equiv  \nu_{n,\ell=0}-(\nu_{n-1,\ell=1} + \nu_{n,\ell=1})/2$). This can be seen in Fig.~\ref{r01r02r13_t6}, which shows the ratio $r_{01} \equiv \delta \nu_{01}(n) / \Delta \nu_{\ell=1}(n)$ and $r_{13} \equiv \delta \nu_{13}(n) / \Delta \nu_{\ell=0}(n+1)$ for rotating and non-rotating models with the same age of 6\,Gyr. Figure~\ref{r01r02r13_t6} also shows that the changes of these asteroseismic ratios increase with the initial velocity of the model. This illustrates that the effects of rotational mixing on the stellar structure and hence on the asteroseismic properties of the models depend on the initial velocity, since a higher initial velocity leads to steeper gradients of the internal angular velocity, resulting in a more effective mixing by shear instability.  

\subsubsection{Models with the same global stellar parameters}

\begin{figure}[htb!]
 \resizebox{\hsize}{!}{\includegraphics{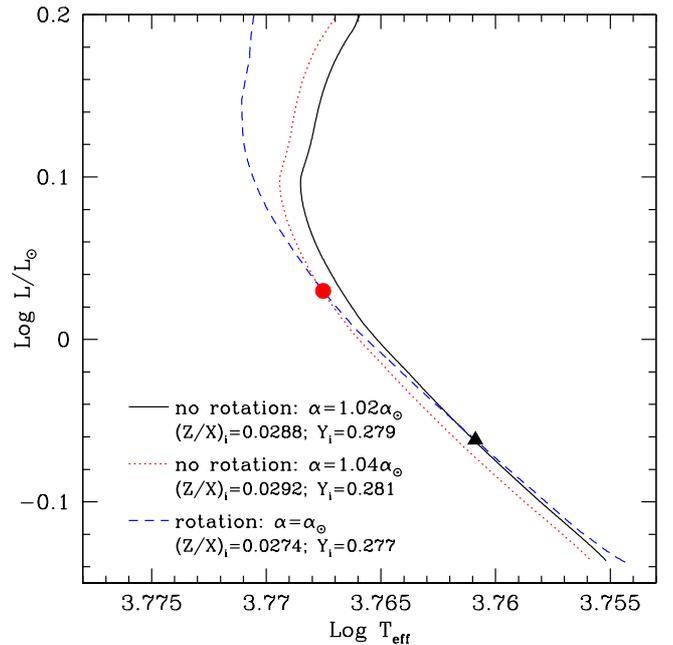}}
  \caption{Evolutionary tracks for 1\,M$_\odot$ models with and without rotation sharing the
same location in the HR diagram. The rotating model (dashed line) is computed with an initial velocity of 50\,km\,s$^{-1}$.
The triangle and the dot indicate the location of the models with an age of 3 and 6\,Gyr, respectively.}
  \label{dhr_ystrot}
\end{figure}

In Sect.~\ref{sec_modini}, the effects of rotational mixing on asteroseismic properties of solar-type stars 
have been studied by comparing rotating and non-rotating models computed with the same initial parameters. 
We are now interested in comparing rotating and non-rotating models sharing 
the same luminosity, effective temperature and surface metallicity, since such a procedure
is closer to observational comparisons. 
Two additional non-rotating stars of 1\,M$_{\odot}$ are then computed in order 
to obtain models with the same location in the HR diagram and surface metallicity as the
rotating 1\,M$_{\odot}$ model studied before.
For this purpose, the initial chemical composition and mixing-length parameter of non-rotating
models are calibrated to reproduce simultaneously the surface metallicity and location in
the HR diagram of the rotating models with an age of 3 and 6\,Gyr. The evolutionary tracks of these models
are shown in Fig.~\ref{dhr_ystrot}. 

The asteroseismic properties of rotating and non-rotating models are first compared by computing
the mean large separation. The rotating model exhibits a mean large separation of 150.1 and 134.4\,$\mu$Hz after
3 and 6\,Gyr, while the corresponding non-rotating models have a mean large separation of
149.9 and 134.5\,$\mu$Hz, respectively. These values are almost identical for rotating and
non-rotating models since the mean large separation is mainly sensitive to the mean density of the star.
For this comparison, the models have the same mass and radius; they exhibit therefore a similar value of the 
mean large separation.  

\begin{figure}[htb!]
 \resizebox{\hsize}{!}{\includegraphics{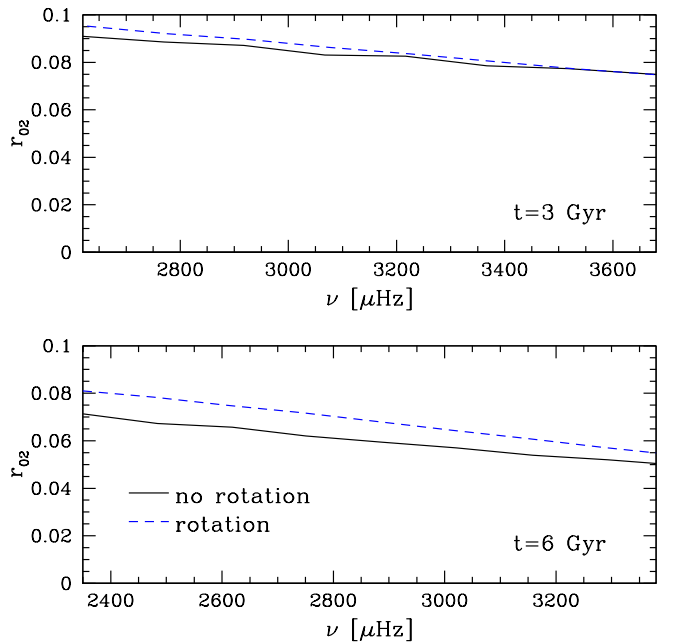}}
  \caption{Variation of the ratio $r_{02}$ with frequency for 1\,M$_\odot$ models with the same global
stellar parameters. The location of these models in the HR diagram is indicated by a triangle and a dot in
Fig.~\ref{dhr_ystrot}.}
  \label{r02_t3t6}
\end{figure}

Concerning the small separation, and more precisely the ratio $r_{02}$ defined above, the situation is quite different.
Figure~\ref{r02_t3t6} indeed shows an increase of the mean value of $r_{02}$ for rotating models compared to
non-rotating models with the same location in the HR diagram and surface metallicity. 
As observed in Sect.~\ref{sec_modini} in models with the same initial parameters, this difference increases during the evolution of the star. The rotating model exhibits indeed a mean value of the ratio
$r_{02}$ of 0.086 and 0.069 after 3 and 6\,Gyr, while the corresponding non-rotating models have a mean ratio of
0.083 and 0.061, respectively. This increase of the mean ratio $r_{02}$ of about 4\% and 13\% after 3 and 6\,Gyr
directly reflects variation of the composition and structure of the central stellar layers
caused by rotational mixing, which become more visible as the age increases. We thus conclude that the increase of the small separation and
the ratio $r_{02}$ discussed above in the case of stellar models computed with the same initial parameters (but sharing
different location in the HR diagram) is still clearly observed for models with the same global stellar parameters.
Figure~\ref{r02_t3t6} shows that this increase is more pronounced at low frequency, leading to a slightly steeper slope in the 
$r_{02}$ versus frequency diagram. Thus, in addition to increasing the mean value of the ratio $r_{02}$,
rotational mixing also changes the frequency dependence of $r_{02}$. This reflects the effects of rotation
on the chemical gradients and in particular the change of the abundance of
hydrogen in the central parts of the star (see Fig.~\ref{profx}). 

By comparing the asteroseismic properties of models that share the same global stellar parameters, 
we conclude that rotational mixing is able to increase the mean
small separation and ratio $r_{02}$ without changing the mean large separation.
These results are particularly interesting in the context of the asteroseismic calibration of stars for which 
solar-like oscillations are observed, since all observational constraints (classical as well as asteroseismic) 
have to be correctly reproduced simultaneously.

\subsection{Models with rotation and magnetic fields}

It is long known that meridional circulation and shear turbulence are actually not sufficient 
to enforce the near uniformity of the solar rotation profile measured by helioseismology. Indeed,
models of solar-type stars including shellular rotation predict an increase in the angular velocity when the distance
to the centre decreases \citep{pin89,cha95,tal97_these,mat98}. This is contradicted
by helioseismic measurements indicating an approximately constant angular velocity
between about 20\% and 70\% of the total solar radius \citep{bro89,kos97,cou03}.
Another process is thus expected to intervene in the transport of angular momentum 
in low-mass stars. Additional clues about the existence and nature of this process come from 
observations of light element abundances in low-mass stars which are efficiently 
spun down via magnetic torquing and lie on the cool side of the lithium dip \citep{tal98,tal03}.

Two main candidates have been proposed to efficiently extract angular momentum from the central
core of a solar-type star, namely magnetic fields \citep[see e.g.][]{mes87,cha93,egg05_mag} 
and travelling internal gravity waves \citep{scha93,kum97,zah97,cha05,tal05,mat08,mat09}. 
Since these processes leave slightly different signatures on the angular rotation profiles, 
it is important to test both of them with asteroseismic techniques.
As of now however only the prescription for angular momentum transport by magnetic fields 
for the Tayler-Spruit dynamo \citep{spr02} is included
in the Geneva stellar evolution code, while the modelling of the transport of angular momentum by internal gravity 
including recent improvements is currently being implemented in this tool.
We thus present here only models computed with both shellular rotation and magnetic fields as prescribed by the Tayler-Spruit dynamo, which
are found to correctly reproduce the helioseismic measurements of the internal rotation of the Sun 
\citep{egg05_mag}. 
This will allow us to investigate the general asteroseismic properties of solar-type stars with quasi-solid body rotation 
of their radiative interior, even if the theoretical prescription for the dynamo \citep[see e.g.][]{den07} as well as its real existence \citep{zah07,gel08}
is still a matter of debate. 
The effects of internal gravity waves on the asteroseismic properties of stellar models as well as 
the influence of other prescriptions for internal magnetic fields 
\citep[e.g.][]{mat05} will be studied in the near future.

\begin{figure}[htb!]
 \resizebox{\hsize}{!}{\includegraphics{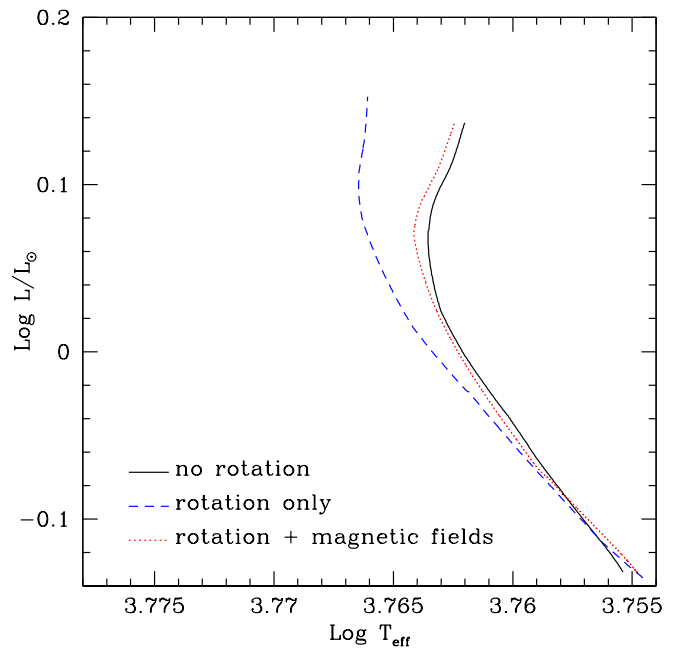}}
  \caption{Evolutionary tracks in the HR diagram for 1\,M$_\odot$ models. 
  The continuous line corresponds to a non-rotating model. The dashed line 
indicates a model computed with rotation only and an initial velocity on the ZAMS
of 50\,km\,s$^{-1}$,
while the dotted line corresponds to a model including rotation and magnetic fields
with an initial velocity of 50\,km\,s$^{-1}$.}
  \label{dhr_magn}
\end{figure}

\begin{figure}[htb!]
 \resizebox{\hsize}{!}{\includegraphics{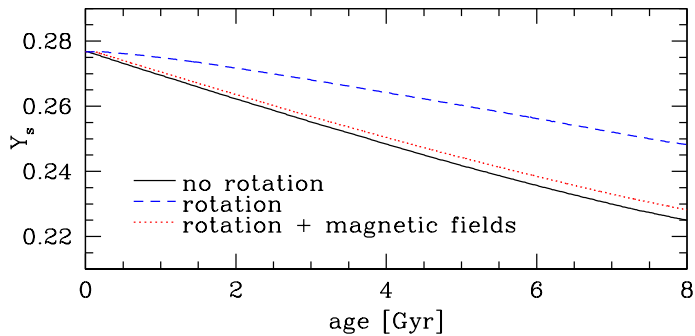}}
  \caption{Surface helium abundance $Y_s$ during the main sequence evolution of models including
rotation only and both rotation and magnetic fields.}
  \label{yst_magn}
\end{figure}

The main-sequence evolution of a 1\,M$_{\odot}$ model including both shellular rotation and the Tayler-Spruit
dynamo is computed for an initial velocity on the ZAMS of 50\,km\,s$^{-1}$. Figure~\ref{dhr_magn} shows the
evolutionary tracks in the HR diagram for this model and for the corresponding models computed
with rotation only and without rotation. All three models include atomic diffusion of helium and heavy elements.
The effects of rotation are strongly reduced when the Tayler-Spruit dynamo is included in the computation. Only a slight shift of the track to
the blue part of the HR diagram is indeed observed for the model with both rotation and magnetic fields,
while the rotating model exhibits a significant increase of the effective temperature. 
These differences observed in the HR diagram can be related to 
changes in the surface chemical composition of the models. 
The variation of the helium surface abundance $Y_{\rm s}$ during the main-sequence is plotted 
in Fig.~\ref{yst_magn} for the three models shown in Fig.~\ref{dhr_magn}. 
As discussed in Sect.~\ref{sec_evo}, rotational mixing counteracts the effects of atomic diffusion 
in the external layers of the star, leading to larger helium surface abundances for rotating models
compared to non-rotating ones. This effect is clearly seen in Fig.~\ref{yst_magn} by comparing the model
with rotation only to the non-rotating model. The model with rotation and magnetic fields exhibits
slightly higher values of the helium surface abundance than the non-rotating model, but lower values than
the model computed with rotation only. We thus see that the efficiency of rotational mixing in the external
layers of solar-type stars is strongly reduced when the effects of the Tayler-Spruit dynamo are taken into account.
As a result, the asteroseismic properties of a model including both rotation and magnetic fields are very similar to the
ones of a non-rotating model. In particular, only a very small increase of about 0.4\,$\mu$Hz of the mean large separation is found near the end of the main sequence (at 8\,Gyr)
for a rotating model computed with the Tayler-Spruit dynamo. 
	
\begin{figure}[htb!]
 \resizebox{\hsize}{!}{\includegraphics{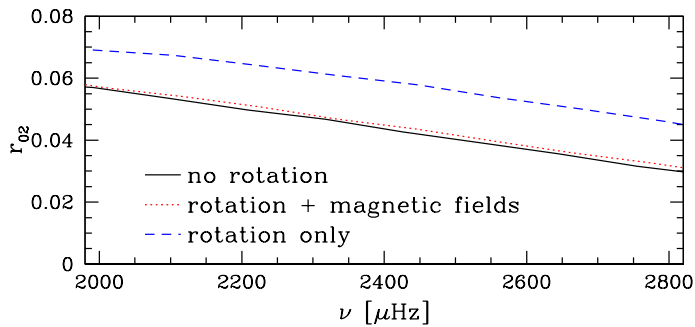}}
  \caption{Variation of the ratio $r_{02}$ with frequency for 1\,M$_\odot$ models with the same age
of 8\,Gyr. The continuous line corresponds to a non-rotating model, while the dashed and dotted lines
correspond to models including rotation only and both rotation and magnetic fields, respectively. 
These rotating models are computed with an initial velocity of 50\,km\,s$^{-1}$.}
  \label{r02_magn_t8}
\end{figure}

The strong reduction of the efficiency of rotational mixing is not limited to the external stellar layers, but is
also observed in the deep interior of solar-type stars. This is clearly seen in Fig.\ref{r02_magn_t8}, which shows 
the ratio $r_{02}$ of the small to large separation as a function of frequency for the three models at the same
age of 8\,Gyr. When both rotation and magnetic fields are included in the computation, a negligible increase
of this ratio is observed compared to the non-rotating model, while the model including only shellular rotation is characterized
by significantly larger values of this ratio.  

That rotational mixing is less efficient in magnetic models of solar-type stars than in models with rotation only is due to the strong decrease of the efficiency of the transport of chemicals by shear mixing when the Tayler-Spruit dynamo is
included in the computation. This decrease is a direct consequence of the near solid body rotation of models including both
rotation and magnetic fields. Compared to the case with rotation only, we also note
a slight increase of the transport of chemicals by meridional circulation for models with magnetic
fields. This greater efficiency of meridional circulation for magnetic models is also due to the near solid body rotation, because uniform rotation creates a strong breakdown of radiative equilibrium. For solar-type
stars, this increase is much smaller though than the strong decrease of the shear turbulent mixing that leads
to a net decrease of the global efficiency of rotational mixing for a rotating model computed with the Tayler-Spruit dynamo. 

The models presented here allow us to illustrate the general asteroseismic properties of solar-type stars with flat rotation profiles 
in their radiative interiors. We wish to recall though that abundance patterns of light elements, and in particular of lithium, constitute 
complementary sensitive indicators of the transport of angular momentum and associated mixing of chemicals in low-mass stars.
In particular, stellar models that correctly account for the internal rotation of the Sun have to 
simultaneously reproduce the so-called lithium dip as well as the temporal evolution of the surface abundance in solar-type stars 
\citep{tal98,tal05,cha05}.
Although it is beyond the scope of the present paper to discuss these aspects in detail, we plan to study quantitatively 
the impact of the Tayler-Spruit dynamo on the light element abundances in the near future.
In particular, it is not clear whether the strong reduction of the efficiency of rotational mixing for magnetic models of low-mass stars
compared to magnetic models of higher mass stars can correctly account for the lithium dip. 

On the other hand it is important to note that stellar models including both shellular rotation and 
the transport of angular momentum by internal gravity waves can simultaneously reproduce the lithium 
behaviour and efficiently extract angular-momentum 
form the central parts of solar-type stars \citep[see][ for more details]{tal02,tal05,cha05}. 
These models therefore provide a promising explanation to obtain a coherent picture of transport
processes in the radiative zone of low-mass stars. 
Because the inclusion of internal gravity waves also results in
a global decrease of the efficiency of rotational mixing, we can expect at first sight a similar impact 
on the asteroseismic properties as for the magnetic models discussed here. It will be particularly interesting however to
investigate this point in detail.  
Work is in progress in this direction.

\section{Conclusions}

The effects of rotational mixing on the properties of solar-type stars have first been studied by comparing stellar models
computed with exactly the same initial parameters except for the inclusion of shellular rotation and atomic diffusion.
The inclusion of rotation is found to change the global properties of a solar-type star with a significant increase of the effective
temperature resulting in a shift of the evolutionary track to the blue part of the HR diagram. These differences in the global properties of rotating
and non-rotating models are related to changes of the chemical composition, because rotational mixing is found to counteract the effects of atomic
diffusion. Consequently, rotating models exhibit larger surface abundances of helium, which lead to a decrease of the opacity in the external layers
and hence to the observed shift to the blue part of the HR diagram. This of course results in a change of the asteroseismic properties of solar-type stars at the same evolutionary stage on the main sequence. Higher values of the large separation are then found for rotating models than for non-rotating ones, because the increase of the effective temperature for rotating models leads to smaller radii and hence to an increase of
the mean density and large separation. In addition to changing the global properties of solar-type stars, rotation is also found to have a significant impact on the structure and chemical composition of the central layers. In particular, rotational mixing brings fresh hydrogen fuel to the central stellar core, thereby enhancing the main-sequence lifetime. At a given age, the central hydrogen mass fraction is larger for rotating models than for models without rotation. The increase of the central hydrogen abundance together with the change of the chemical profiles in the central layers due to rotational mixing result in a significant increase of the values of the small separations and of the ratio of small to large separations.  

After comparing models computed with the same initial parameters, the effects of rotational mixing on the asteroseismic properties of models sharing the same global stellar parameters, and in particular the same location in the HR diagram, are investigated. The values of the large separation are then almost identical for these rotating and non-rotating models since they have the same mass and radius and hence the same mean density. However, the effects of rotational mixing on the central layers can be clearly revealed by asteroseismic observations, since a significant increase of the small separations and of the ratio of small to large separations is found for rotating models compared to non-rotating models with the same location in the HR diagram. We thus note that rotational mixing is able to increase the mean small separations and ratio of small to large separations without changing the value of the mean large separation. This result is particularly interesting in the context of the asteroseismic calibration of a solar-type star, for which all observational constraints (classical as well as asteroseismic) have to be correctly reproduced simultaneously. 

The effects of internal magnetic fields in the framework of a dynamo that possibly occurs in the radiative zone are also studied by computing rotating models of solar-type stars including the Tayler-Spruit dynamo. We find that the efficiency of rotational mixing is strongly reduced when the effects of the Tayler-Spruit dynamo are taken into account. The asteroseismic properties of a solar-type star model including both rotation and magnetic fields are then very similar to the ones of a non-rotating model. Indeed, only a very small increase of the values of the large and small separation is found for a rotating model computed with the Tayler-Spruit dynamo compared to a non-rotating model. This is because the strong decrease of the efficiency of shear mixing is not compensated by the limited increase of the transport of chemicals by meridional circulation when internal magnetic fields are taken into account. It is interesting to recall here that the situation is quite different for magnetic models of massive stars. For a massive star computed with the Tayler-Spruit dynamo, the strong breakdown of radiative equilibrium imposed by the approximately flat rotation profile leads indeed to an increase of the efficiency of meridional circulation which is larger than the decrease of the shear turbulent mixing \citep{mae05}. 
The magnetic models presented here allow us to investigate the general asteroseismic properties of low-mass stars with radiative interiors rotating as quasi-solid bodies. It is important to recall however that alternative processes such as those induced by internal gravity waves can also ensure a very efficient angular momentum transport. The asteroseismic characteristics of the corresponding models will have to be investigated in the near future.

We finally conclude that asteroseismic observations are able to reveal the changes in the internal structure and global properties of solar-type stars induced by rotational mixing. In particular, the effects of rotational mixing on the chemical composition of the central layers of a solar-type star significantly changes its asteroseismic properties. We thus see here an interesting complementarity between spectroscopic and asteroseismic observations, since surface abundances can be determined
by spectroscopic measurements while asteroseismic data can give us valuable insights into the chemical composition of the central core. This is exactly what is needed to progress in our understanding of the transport processes at work in stellar interiors. In addition to the indirect effects of rotation on the asteroseismic properties of a star through the change of its internal structure by rotational mixing, direct effects of rotation related to the breaking of the spherical symmetry can also be revealed by asteroseismic measurements of rotational splittings. New asteroseismic data from ground-based observations and space missions like CoRoT and Kepler are thus particularly valuable in order to provide us with new insight into the internal rotation of solar-type stars and with constraints on how to model it best.

\begin{acknowledgements}
We would like to thank J. Christensen--Dalsgaard for providing us with the Aarhus adiabatic pulsation code.
Part of this work was supported by the Swiss National Science Foundation.
\end{acknowledgements}

\bibliographystyle{aa} 
\bibliography{biblio} 

\end{document}